\documentclass[fleqn,usenatbib]{mnras}

\usepackage[T1]{fontenc}
\usepackage{multirow}
\usepackage{array}
\usepackage{threeparttable}
\usepackage{longtable}
\usepackage{lscape}
\usepackage{pdflscape}
\DeclareRobustCommand{\VAN}[3]{#2}
\let\VANthebibliography\thebibliography
\def\thebibliography{\DeclareRobustCommand{\VAN}[3]{##3}\VANthebibliography}

\usepackage{graphicx}	
\usepackage{amsmath}	
\usepackage{amssymb}	
\usepackage{txfonts}
\usepackage{mathrsfs}
\usepackage{multicol}
\usepackage{color}

\title[A Statistical Analysis of Hubble Tension]{The Hubble Tension Survey: A Statistical Analysis of the 2012-2022 Measurements}

\author[Wang et al.]{
Bao Wang$^{1,2}$,
Mart\'{i}n L\'{o}pez-Corredoira$^{3,4,5}$ \thanks{E-mail: martin@lopez-corredoira.com}
and Jun-Jie Wei$^{1,2}$ \thanks{E-mail: jjwei@pmo.ac.cn}
\\
$^{1}$Purple Mountain Observatory, Chinese Academy of Sciences, Nanjing 210023, China\\
$^{2}$School of Astronomy and Space Sciences, University of Science and Technology of China, Hefei 230026, China\\
$^{3}$Instituto de Astrof\'{i}sica de Canarias, E-38205 La Laguna, Tenerife, Spain\\
$^{4}$PIFI-Visiting Scientist 2023 of China Academy of Sciences at Purple Mountain Observatory, Nanjing 210023 \\ and National Astronomical Observatories, Beijing 100012, China\\
$^{5}$Departamento de Astrof\'{i}sica, Universidad de La Laguna, E-38206 La Laguna, Tenerife, Spain}

\pubyear{2023}

\begin{document}
\label{firstpage}
\pagerange{\pageref{firstpage}--\pageref{lastpage}}
\maketitle

\begin{abstract}
In order to investigate the potential Hubble tension, we compile a catalogue of 216 measurements of the Hubble--Lema\^itre constant $H_0$ between 2012 and 2022, which includes 109 model-independent measurements and 107 $\Lambda$CDM model-based measurements.
Statistical analyses of these measurements show that
the deviations of the results with respect to the average $H_0$ are far larger than expected from their error
bars if they follow a Gaussian distribution. We find that $x\sigma $ deviation is indeed equivalent in a Gaussian
distribution to $x_{\rm eq}\sigma$ deviation in the frequency of values,
where $x_{\rm eq}=0.72x^{0.88}$.
Hence, a tension of 5$\sigma$, estimated between the Cepheid-calibrated type Ia supernovae and cosmic microwave background (CMB) data, 
is indeed a 3$\sigma$ tension in equivalent terms of a Gaussian distribution of frequencies.
However, this recalibration should be independent of the data whose tension we
want to test. If we adopt the previous analysis of data of 1976-2019, the equivalent tension is
reduced to $2.25\sigma$.
Covariance terms due to correlations of measurements do not significantly change the results.
Nonetheless, the separation of the data into two blocks with $H_0<71$ and $H_0\ge 71$ km s$^{-1}$ Mpc$^{-1}$ finds compatibility with a Gaussian distribution for each of them without removing any outlier.
These statistical results indicate that the underestimation of error bars for $H_0$ remains prevalent over the past decade, dominated by systematic errors in the methodologies of CMB and local distance ladder analyses.
\end{abstract}




\section{Introduction}
Few problems in astrophysics have received as much attention in the last years as what is called `Hubble tension'.
Hundreds or thousands of papers dedicated to investigating the observations that originate the tension within the standard cosmological model or to propose alternative scenarios have been produced (see reviews by \citealt{DiV21,Per22,Abd22,Hu23,Vagnozzi_2023}).
The tension was mainly triggered with the claim in 2019 of a Hubble--Lema\^itre constant $H_0$ estimated from the local
Cepheid--type Ia supernova (SN Ia) distance ladder being at odds with the value extrapolated from Cosmic Microwave Background (CMB) data, assuming the standard $\Lambda$CDM cosmological model, $74.0\pm 1.4$ \citep{Rie19} and $67.4\pm 0.5$ km s$^{-1}$ Mpc$^{-1}$ \citep{Planck2020}, respectively, which gave an incompatibility at the 4.4$\sigma$ level. This tension was later increased up to 6$\sigma $ depending on the datasets considered \citep{DiV21}. Very recently, the latest result from the Cepheid--SN Ia sample is $H_{0}=73.04\pm 1.04$ km s$^{-1}$ Mpc$^{-1}$ \citep{Riess2022}, representing a 5$\sigma$ tension with that
estimated from CMB data.

This tension should not be so surprising, given the number of systematic errors that may arise in the measurements.
As a matter of fact, there have always been tensions between different measurements in the values of the Hubble--Lema\^itre constant, which has not received so much attention. Before the 1970s, due to different corrections of errors in the calibration
of standard candles, the parameter continuously decreased its value, making incompatible measurements of different epochs \citep{Tul23}. But even after the 1970s, a tension has always been present. The compilations of values until the beginning of the 2000s showed
an error distribution that was strongly non-Gaussian, with significantly larger probability in the tails of the distribution than predicted by a Gaussian distribution: the 95.4\% confidence-level (CL) limits are 7.0$\sigma $ in terms of the quoted errors \citep{Che03}.
The nature of the possible systematic errors is unknown, and they may dominate over the statistical errors.
Twenty years ago, it was estimated that these systematic errors might be of the order of 5 km s$^{-1}$ Mpc$^{-1}$ (95\% CL) \citep{Got01}.

The common likelihood functions used by astronomers contain the assumption of Gaussian errors \citep{D'Agostini2005}, which is also a requirement of the central limit theorem.
Statistical analyses of the measurements of the Hubble--Lema\^itre constant $H_0$ between 1976 and 2019 \citep{Faerber2020,Martin2022} have also shown that the dispersion of its value is far much larger than what would be expected in a Gaussian distribution given the published error bars.
The only solution to understand this dispersion of values is assuming that most of the statistical error bars associated with the observed parameter measurements have been underestimated, or the systematic errors were not properly taken into account.
The fact that the underestimation of error bars for $H_0$ is so common might explain the apparent discrepancy of values.
Indeed, a recalibration of the probabilities with this sample of measurements to make it compatible with a Gaussian distribution of deviations finds that a tension of 4.4$\sigma $ would be indeed a 2.1$\sigma $ tension in equivalent terms of a Gaussian distribution of frequencies, and a tension of 6.0$\sigma $ would be indeed a 2.5$\sigma $ tension in equivalent terms of a Gaussian distribution of frequencies \citep{Martin2022}. That is, we should not be surprised to find those 4-6 $\sigma $ tensions,
because they are much more frequent than indicated by the Gaussian statistics, and they stem from underestimation of errors,
not from real tensions in the background of physics or cosmology.

In this paper, we want to extend this type of historical statistical analyses focusing only in the years 2012-2022.
We know the statistical and systematic errors are much smaller now than some decades ago, but it is still worth to check whether
the distribution of these errors follows what is expected in a Gaussian distribution.
Clearly, if we focus on the Hubble tension between SN Ia data and CMB data, the answer is negative,
but apart from these two types of sources, we want to explore other measurements too and globally evaluate the distribution.

In Sect. \ref{Bibliographical}, we give a description of the bibliographical data we used for our statistical analyses and
the criteria to select them. Statistical analyses are presented in Sect. \ref{Statistical}. Recalibration of data in order to comply with a Gaussian distribution is presented in Sect. \ref{recalibration}. We are aware that many of the data are correlated, they
were measured by the same teams and/or with similar calibrators or methods, so we add in Sect. \ref{Sec:5} some considerations
on the effect of removing correlated data. The last section summarizes and discusses the results and the
possible origin of the underestimated statistical errors and/or `unknown' systematic errors.


\section{Bibliographical Data}
\label{Bibliographical}

\begin{figure}
	\centering
	\includegraphics[width=0.48\textwidth]{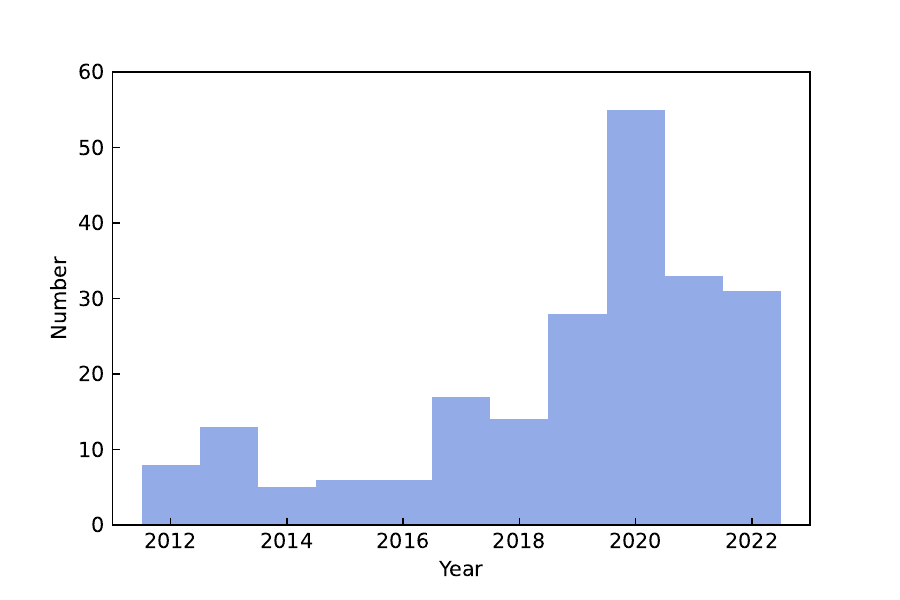}
	\includegraphics[width=0.48\textwidth]{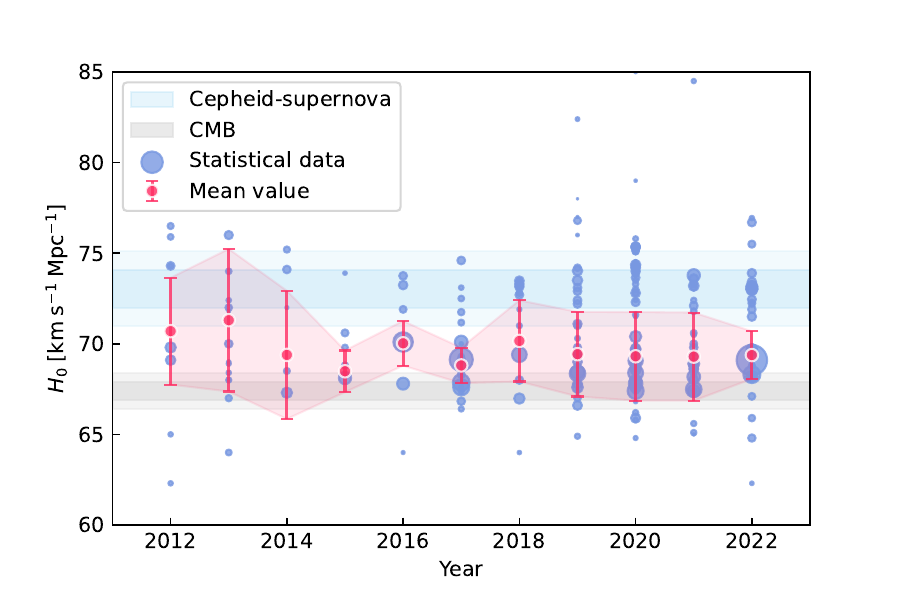}
	\caption{Histogram and scatter diagram of 216 measurements of the Hubble--Lema\^itre constant $H_0$ between 2012 and 2022.
	The size of the blue scatter points is inversely proportional to the error.
	Obviously, there is potential for increasingly precise estimation of $H_0$.
 The red points are weighted averages with weighted standard deviations for each year.
	The results derived from the local Cepheid-supernova distance ladder and CMB data are also included for comparison.
	\label{fig:year}
	}
\end{figure}

We conduct a comprehensive search for $H_0$ measurements between the years 2012 and 2022 in the NASA Astrophysics Data System\footnote{\href{https://ui.adsabs.harvard.edu/}{https://ui.adsabs.harvard.edu/.}}, adhering to the following selection criteria:
\begin{enumerate}
	\item The $H_0$ values were reported by published papers.
	\item We only pay attention to the $H_0$ measurements inferred from model-independent methods and within the context of standard $\Lambda$CDM. 
 Note that hundreds or thousands of papers  dedicated to investigating the Hubble tension. A substantial fraction of them focuses on proposing alternative cosmological models for measuring $H_0$ in order to narrow the discrepancy between the CMB and SN Ia observations. Those $H_0$ measurements obtained under models other than $\Lambda$CDM\footnote{Numerous models proposed for solving the $H_0$
tension are divided into 11 major categories with 123 subcategories by \cite{DiV21}.} are not included in our sample, since they potentially interfere with the analysis of the Hubble tension.
\end{enumerate}
Based on the aforementioned criteria, we compile a total of 216 $H_0$ measured values, with 109 values measured from model-independent methods and 107 values obtained under the $\Lambda$CDM model.
For the complete catalogue, please refer to \autoref{TabB1} provided in the Appendix.
The histogram and scatter of these data are plotted in \autoref{fig:year}.
The size of the scatter dots is inversely proportional to the error.
We can find a significant increase in numbers of the measured $H_0$ after the year 2019, primarily attributed to the renowned Hubble tension problem, while concurrently witnessing an improvement in measurement accuracy.
It is essential to reexamine these $H_0$ measurements from a statistical perspective.
Therefore, we employ the statistical analysis approach proposed by \cite{Martin2022} to investigate the $H_0$ tension over the past 11 years.

\section{Statistical Analysis}
\label{Statistical}

\subsection{Distributions of Statistical Data}
\label{subc.Distribution}

\begin{figure}
	\centering
	\includegraphics[width=0.48\textwidth]{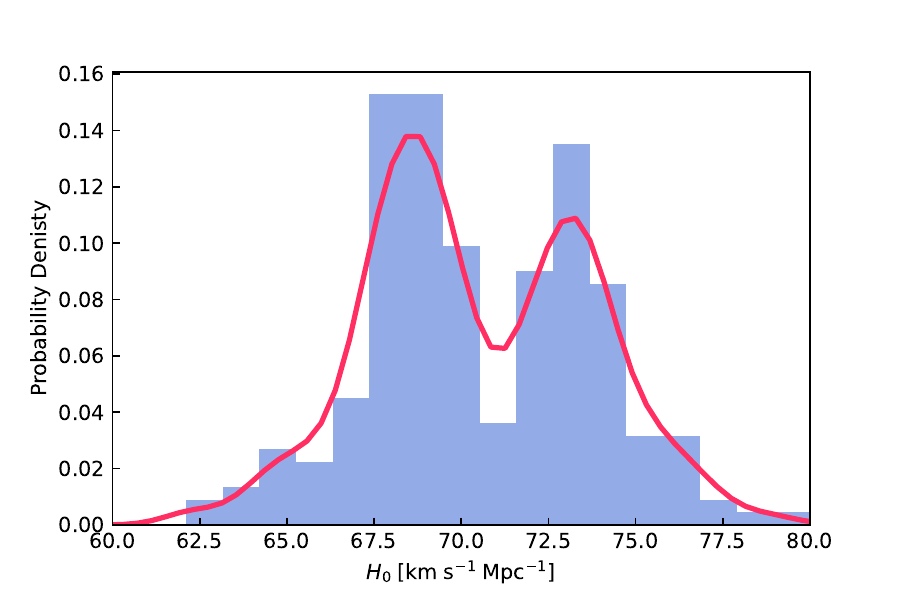}
	\includegraphics[width=0.48\textwidth]{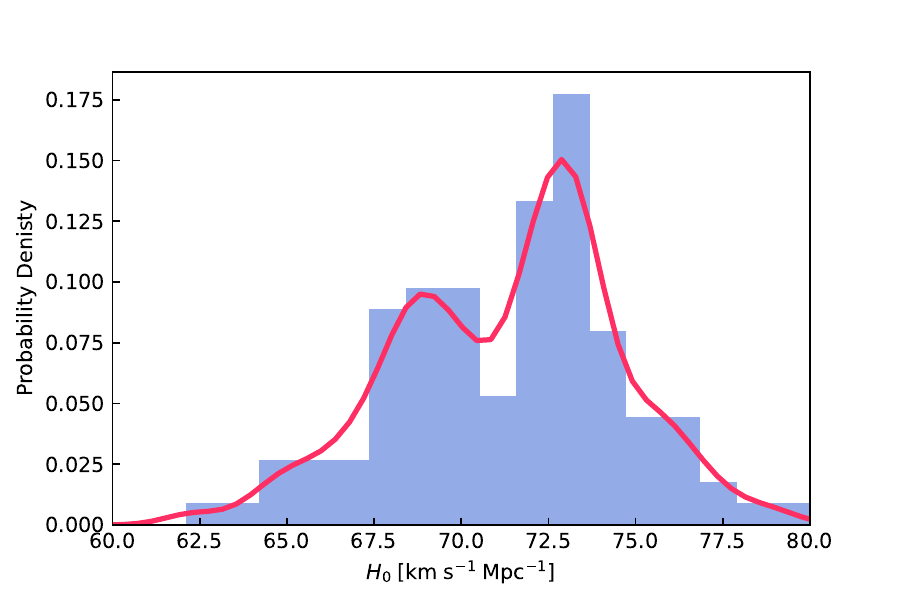}
	\includegraphics[width=0.48\textwidth]{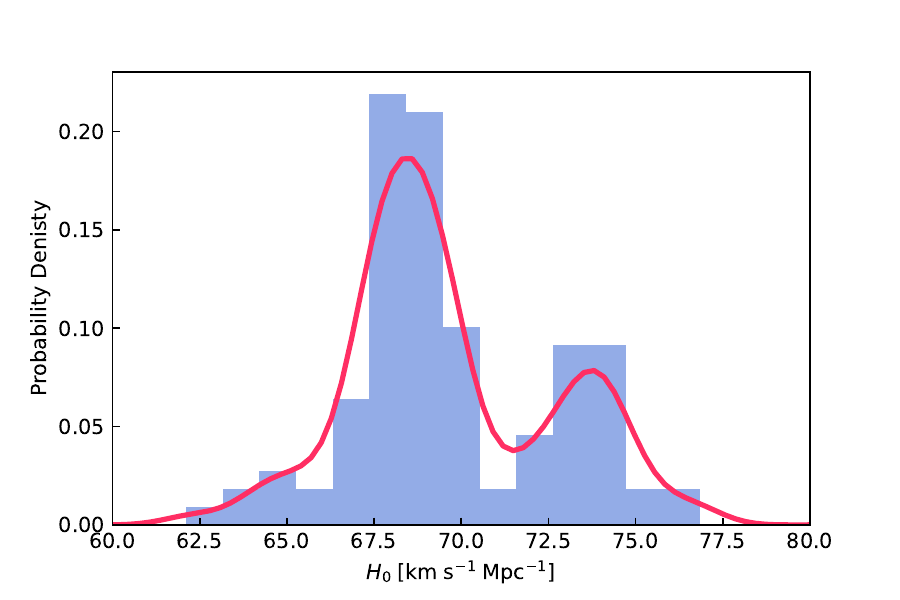}
	\caption{Histograms of three categories:
	(top panel) 216 complete measurements, (middle panel) 109 model-independent measurements and (bottom panel) 107 $\Lambda$CDM model-based measurements.
	Kernel density estimation curves are depicted as red lines in each histogram.
	For clarity, several extreme values exceeding 80 km s$^{-1}$ Mpc$^{-1}$ are not displayed.
		\label{fig:Hist}
	}
\end{figure}

The statistical data of $H_0$ are divided into three categories: complete, model-independent and $\Lambda$CDM
model-based measurements.
These categorizations can be distinguished based on the statements in the respective articles (see \autoref{TabB1}).
To visualize the distributions of these measurements, in \autoref{fig:Hist} we plot histograms for each category.
Notably, significant bimodal distributions are observed in all three subplots, which imply the possible tension: the measurements are clustered around $67.4 \pm 0.5 \; {\rm km\;s^{-1}\;Mpc^{-1}}$ from the CMB data \citep{Planck2020}
and $73.04 \pm 1.04 \; {\rm km\;s^{-1}\;Mpc^{-1}}$ from the local Cepheid-supernova distance ladder \citep{Riess2022}.
Despite the model-independent measurements tending to favor the results of local distance ladders and the $\Lambda$CDM model-based measurements tending to favor the results of Planck CMB observations, the bimodal distributions still exist in both cases.
This phenomenon has not been reported in previous analyses conducted using $H_0$ statistical data \citep{Chen2011, Croft2015, Zhang2018, Faerber2020, Martin2022}.

Moreover, from the perspective of early and late Universe observations, our statistics also reveal some intriguing situations. The methods of the early Universe observations, mainly including CMB and baryon acoustic oscillation (BAO) data, consistently have results around $68 \; {\rm km\;s^{-1}\;Mpc^{-1}}$ in $\Lambda$CDM model-based measurements, aligning with expectations. However, it was previously thought that the $H_0$ measured in late Universe was approximately $73 \; {\rm km\;s^{-1}\;Mpc^{-1}}$, which does not match our statistics. Almost all of the model-independent measurements in our statistics are derived from the late Universe observations, but a considerable proportion of these measurements deviate from the value of $73 \; {\rm km\;s^{-1}\;Mpc^{-1}}$.

\subsection{Statistical Significance of the Bimodality}
\label{subc.71}

In this subsection, our focus is on assessing the statistical significance of the double peaks in the distribution.
To achieve this objective, we employ the dip test, a general method for testing multimodality of distributions  \citep{hartigan1985}.
The dip test quantifies multimodality in a sample by calculating the  discrepancy between the empirical distribution function of
 the samples and the unimodal distribution function that minimizes the maximum discrepancy.
We implement this approach by using the $diptest$\footnote{\href{https://github.com/RUrlus/diptest/blob/stable}{https://github.com/RUrlus/diptest/blob/stable.}} package in Python.
The $p$ values for the dip test range between 0 and 1, representing the probability of  unimodality.
The $p$ values less than 0.05 indicate significant multimodality.

In addition to the three categories mentioned in the previous subsection, we further segment the complete $H_0$ measurements into two blocks: $H_0<71\;{\rm km\;s^{-1}\;Mpc^{-1}}$ (118 data) and $H_0\ge 71\;{\rm km\;s^{-1}\;Mpc^{-1}}$ (98 data) for comparison.
The results are summarized in the last column of \autoref{Tab1}.
The complete measurements result in $p=0.01$ and the measurements for the two blocks result in  $p=0.96$ and $p=0.98$, respectively.
This reveals that the measurements of two blocks exhibit unimodality ($p \gg 0.05$), whereas complete measurements exhibit significant multimodality ($p=0.01 < 0.05$), specifically bimodality.
This implies a potential $H_0$ tension.
For the model-independent and $\Lambda$CDM model-based measurements, the results of the dip test are $p=0.46$ and $p=0.16$, respectively, indicating that the evidence of multimodality is not robust, and the unimodality is not statistically 
 significant either.

\subsection{Statistical Significance of model-independent and $\Lambda$CDM model-based measurements}

\begin{table*}
	\centering
	\caption{The results of dip tests, weighted averages and $\chi^2$ fittings in three categories: complete, model-independent and $\Lambda$CDM model-based measurements.
    The $Q$ values measure the statistical significance for $\chi^2$ fittings, and the $p$ values measure the significance of the unimodality for dip tests.}
	\label{Tab1}
	\renewcommand{\arraystretch}{1.2}
	\begin{tabular}{cccccc} 
		\hline
		 & Number & $\overline{H_{0}} \; (\sigma)$ & $\chi^2$ & $Q$ & $p$ \\
		 & & $({\rm km\;s^{-1}\;Mpc^{-1}})$ & & \\
		\hline
		Complete & 216 & $69.35\pm0.12$ & 515.99 & $8.85 \times 10^{-27}$ & 0.01 \\
		Model-independent & 109 & $70.82 \pm 0.22$ & 181.48 & $1.23 \times 10^{-5}$ & 0.46 \\
		$\Lambda$CDM model-based & 107 & $68.94 \pm 0.13$ & 237.56 & $4.36 \times 10^{-12}$ & 0.16 \\
		\hline
  		$H_0<71 \; {\rm km\;s^{-1}\;Mpc^{-1}}$ & 118 & $68.78 \pm 0.07$ & 95.09 & 0.93 & 0.96 \\
		$H_0\ge 71 \; {\rm km\;s^{-1}\;Mpc^{-1}}$ & 98 & $73.53 \pm 0.13$ & 30.05 & $\sim 1.00$ & 0.98 \\
		\hline
	\end{tabular}
\end{table*}

To investigate the causes of the bimodal distributions, we continue to analyse the statistical significance of three categories: complete, model-independent and $\Lambda$CDM model-based measurements.
We calculate the weighted average value $\overline{H_{0}}$ to analyze the statistical characteristic, whereby each data value is assigned a weight based on its inverse variance.
Then we can obtain $\chi^2$ for $\overline{H_{0}}$ of three categories:
\begin{equation}\label{chi2}
	\chi^2=\sum^{N}_{i=1}{\frac{\left(H_{0,i}-\overline {H_{0}}  \right)^2}{\sigma_i^2}} ,
\end{equation}
where $H_{0,i}$ and $\sigma_i$ are the measured value and standard deviation, respectively.
The weighted average values $\overline{H_{0}}$, along with their corresponding standard deviations $\sigma$, $\chi^2$ and $Q$ values are summarized in \autoref{Tab1}.
The index $Q$ is used to measure their statistical significance, representing the probability that sample differences arise solely from chance errors.
Typically, the statistical significance is satisfied when $Q>0.05$.
The three categories yield weighted averages of $\overline{H_{0}}=69.35\pm 0.12$ ${\rm km\;s^{-1}\;Mpc^{-1}}$ ($Q=8.85\times 10^{-27}$), $\overline{H_{0}}=70.82\pm 0.22$ ${\rm km\;s^{-1}\;Mpc^{-1}}$ ($Q=1.23\times 10^{-5}$), and $\overline{H_{0}}=68.94\pm 0.13$ ${\rm km\;s^{-1}\;Mpc^{-1}}$ ($Q=4.36\times 10^{-12}$), respectively.
The $\chi^2$ fittings of two $H_0$ blocks are also displayed, with $Q \sim 1$.
However, for all three categories, the $Q$ values are much lower than the threshold 0.05, indicating that it is highly unlikely that the observed trends are due to chance errors.

It is important to note that the implicit assumption in our calculations is that the covariance of each $H_0$ measurement
is ignored. A considerable fraction of the measurements were not independent at all, since there were many duplicate data being used. If the data are treated as independent variables, a conservative limit of the real dispersion of data would be
obtained. This is because the `effective' number of degrees of freedom is smaller than the number assuming independence,
leading to an increase of the reduced $\chi^2$ value.
These defects also make the cause of the bimodal distributions uncertain, as they may arise from a failure to eliminate the correlated data.
More discussion can refer to \autoref{Sec:5}.

\subsection{Outliers Screening}

\begin{table*}
	\centering
	\caption{The results of weighted averages and $\chi^2$ fittings in three categories after removing the outliers.
	The number of outliers and minimal deviations are also displayed.}
	\label{Tab2}
	\renewcommand{\arraystretch}{1.2}
	\begin{tabular}{ccccccc}
		\hline
		 & $x_{\rm min}$ & Outliers & Number & $\overline{H_{0}} \; (\sigma)$ & $\chi^2$ & $Q$ \\
		 & & & & $({\rm km\;s^{-1}\;Mpc^{-1}})$ & & \\
		\hline
		Complete & 2.4 & 27 & 189 & $69.17 \pm 0.09$ & 216.42 & 0.08  \\
		Model-independent & 3.6 &1 & 108 & $72.45 \pm 0.21$ & 84.20 & 0.95 \\
		$\Lambda$CDM model-based & 2.6 & 13 & 94 & $68.85 \pm 0.10$ & 106.33 & 0.16 \\
		\hline
	\end{tabular}
\end{table*}

Removing sufficient outliers which deviate significantly from the mean would result in $Q$ values exceeding 0.05.
It is imperative to investigate how many outliers lead to a loss of statistical significance.
Prior to removal, we define the number of $\sigma$ deviations between the measurements $H_{0,i}$ and the average $\overline{H_{0}}$ as
\begin{equation}\label{xsigma}
	x=\frac{\left|H_{0,i}-\overline{H_{0}} \right|}{\sigma_i}.
\end{equation}
To ensure a significance level of $Q>0.05$, data with $x>x_{\rm min}$ are excluded, and the calculations for $\overline{H_{0}}$ and $\chi^2$ follow the methodology described in the previous subsection.
Our results are listed in \autoref{Tab2}, where the weighted averages $\overline{H_{0}}$ are $69.17\pm 0.09$, $72.45\pm 0.21$ and $68.85\pm 0.10$ ${\rm km\;s^{-1}\;Mpc^{-1}}$, respectively.
Additionally, the 27 outliers of the complete category are listed in \autoref{Tab3}.
Compared to the values obtained in \autoref{Tab1},  several pieces of information can be inferred:
\begin{enumerate}
	\item The $H_0$ measurements from model-independent methods perform best.
	The number of outliers is only one, with a value of $69.13\pm 0.24$ ${\rm km\;s^{-1}\;Mpc^{-1}}$, obtained from combining Hubble parameter $H(z)$, Sunyaev-Zel'dovich effect and X-ray data \citep{Huang2017}.
	The outlier is due to its remarkably low error.
	\item The outliers in the complete dataset predominantly arise from the observations of Cepheids+SNe Ia, lensing, CMB and BAO. Due to their small error ranges, the deviations are obvious, especially for the renowned results from \cite{Riess2022} and \cite{Planck2020}.
	And most of the outliers were obtained after the year 2019.
	\item The outliers in the Model-independent and $\Lambda$CDM model-based dataset are less than in the complete one, which implies the possible tension between them.
\end{enumerate}

Based on the available evidence, there are signs of the Hubble tension. However, the degree of tension may be overestimated. Therefore, our subsequent analysis aimed to quantify this potential overestimation.

\begin{table*}
	\centering
	\caption{27 outliers: measurements of $H_0$ in which $| H_0-\overline{H_{0}} | /\sigma>2.4$, where $\overline{H_{0}} = 69.35\; {\rm km\;s^{-1}\;Mpc^{-1}}$ is the weighted average of the 216 measurements of the literature.}
	\label{Tab3}
	\renewcommand{\arraystretch}{1.2}
	\begin{tabular}{ccccc} 
		\hline
		Year & $H_0$ & $\left| H_0-\overline{H_{0}} \right| /\sigma$ & Authors & Methods \\
		& $({\rm km\;s^{-1}\;Mpc^{-1}})$ & & &\\
		\hline
		2013 & $76.00 \pm 1.90$ & 3.50 & Fiorentino et al. & Cepheids+SNe Ia   \\
		2017 & $67.60\pm 0.50$ & 3.49 & Alam et al. & BAO+SNe Ia\\
		2017 & $67.87 \pm 0.46$ & 3.21 & Chavanis et al. & Planck 2015+Lensing+BAO+JLA+HST    \\
		2017 & $74.60 \pm 2.10$ & 2.50 & Wang et al. & Angular diameter distance    \\
		2018 & $73.48 \pm 1.66$ & 2.49 & Riess et al. & Cepheids+SNe Ia   \\
		2019 & $76.80 \pm 2.60$ & 2.87 & Chen et al. & Gravitational lensing   \\
		2019 & $73.50\pm 1.40$ & 2.97 & Reid et al. & Cepheids+SNe Ia  \\
		2019 & $74.03 \pm 1.42$ & 3.30 & Riess et al. & Cepheids+SNe Ia   \\
		2020 & $74.36 \pm 1.42$ & 3.53 & Camarena et al. &  SNe Ia+Angular BAO+$r_{\rm CMB}$ prior  \\
		2020 & $75.32 \pm 1.68$ & 3.56 & Camarena et al. & SNe Ia+Anisotropic BAO+$M_B$ prior   \\
		2020 & $75.35\pm 1.68$   & 3.58 & Camarena et al.  & SNe Ia+$M_B$ prior \\
		2020 & $75.36 \pm 1.68$ & 3.58 & Camarena et al. & SNe Ia+Angular BAO+$M_B$ prior   \\
		2020 & $74.00 \pm 1.75$ & 2.66 & Millon et al. & Gravitational lensing   \\
		2020 & $74.20 \pm 1.60$ & 3.03 & Millon et al. & Gravitational lensing   \\
		2020 & $67.40 \pm 0.50$ & 3.89 & Planck Collaboration et al. & Planck2018 \\
		2020 & $73.60 \pm 1.70$ & 2.50 & Qi et al. & Gravitational lensing   \\
		2020 & $74.30 \pm 1.90$ & 2.61 & Wei et al. &  Lensing+SNe Ia  \\
		2021 & $67.49 \pm 0.53$ & 3.50 & Balkenhol et al. & Planck2018+SPT+ACT \\
		2021 & $73.78 \pm 0.84$ & 5.28 & Bonilla et al. & SNe Ia+CC+BAO+H0LiCOW \\
		2021 & $73.20 \pm 1.30$ & 2.97 & Riess et al. & Cepheids+SNe Ia   \\
		2022 & $75.50 \pm 2.50$ & 2.46 & Kourkchi et al. & Tully-Fisher Relation  \\
		2022 & $73.90 \pm 1.80$ & 2.53 & M\"{o}rtsell et al. & Cepheids+SNe Ia   \\
		2022 & $73.20 \pm 1.30$ & 2.97 & M\"{o}rtsell et al. & Cepheids+SNe Ia   \\
		2022 & $76.70 \pm 2.00$ & 3.68 & M\"{o}rtsell et al. & Cepheids  \\
		2022 & $73.04 \pm 1.04$ & 3.55 & Riess et al. & Cepheids+SNe Ia   \\
		2022 & $73.01 \pm 0.99$ & 3.70 & Riess et al. & Cepheids+SNe Ia   \\
		2022 & $73.15 \pm 0.97$ & 3.92 & Riess et al. & Cepheids+SNe Ia   \\
		\hline
	\end{tabular}
\end{table*}


\section{Recalibration of Probabilities}\label{Sec:Recalibration}
\label{recalibration}

\begin{figure}
	\centering
	\includegraphics[width=0.49\textwidth]{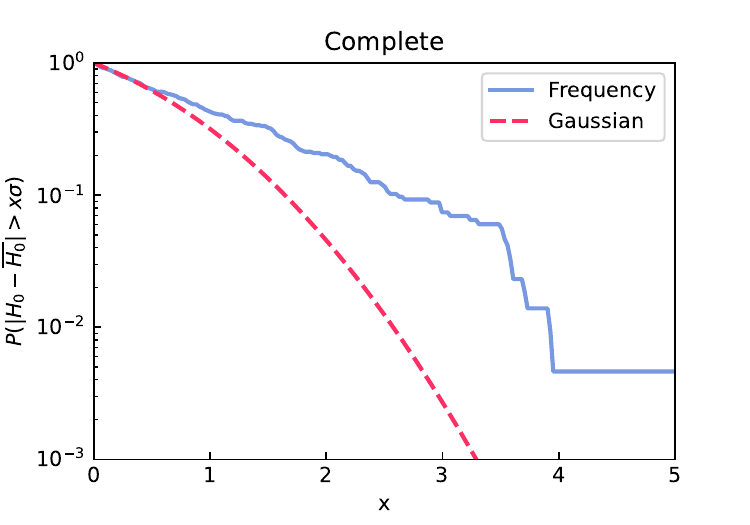}
	\includegraphics[width=0.49\textwidth]{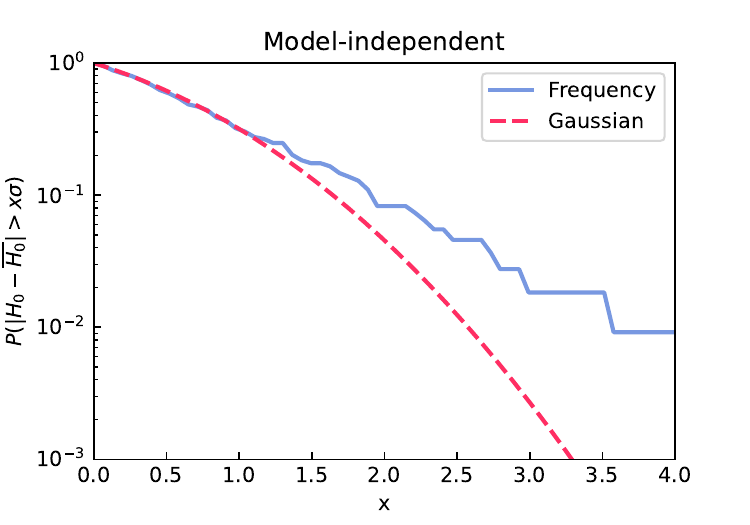}
	\includegraphics[width=0.49\textwidth]{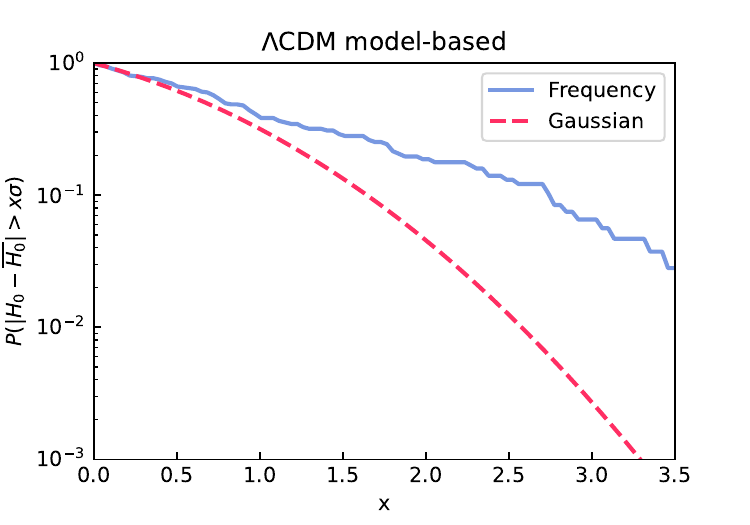}
	\caption{Probabilities of deviation larger than $x \sigma$ for three categories:
	(top panel) 216 complete measurements, (middle panel) 109 model-independent measurements and (bottom panel) 107 $\Lambda$CDM model-based measurements.
	The red dashed lines denote the expected probabilities if the errors are Gaussian.
		\label{fig:Prob}
	}
\end{figure}

\begin{figure}
	\centering
	\includegraphics[width=0.49\textwidth]{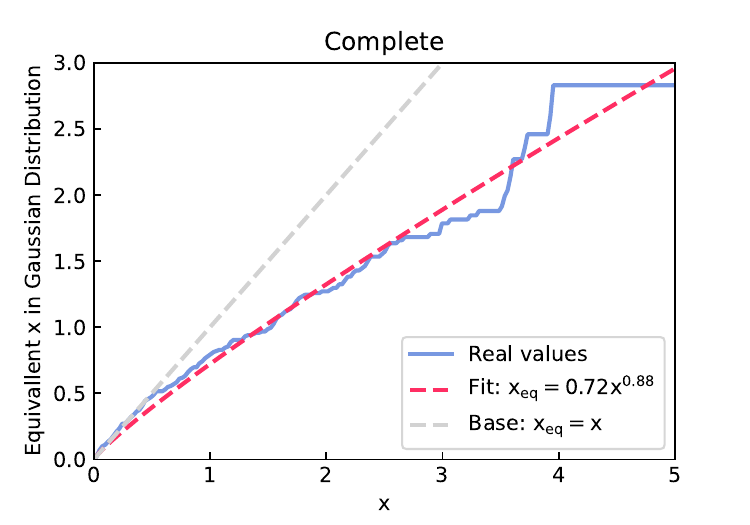}
	\includegraphics[width=0.49\textwidth]{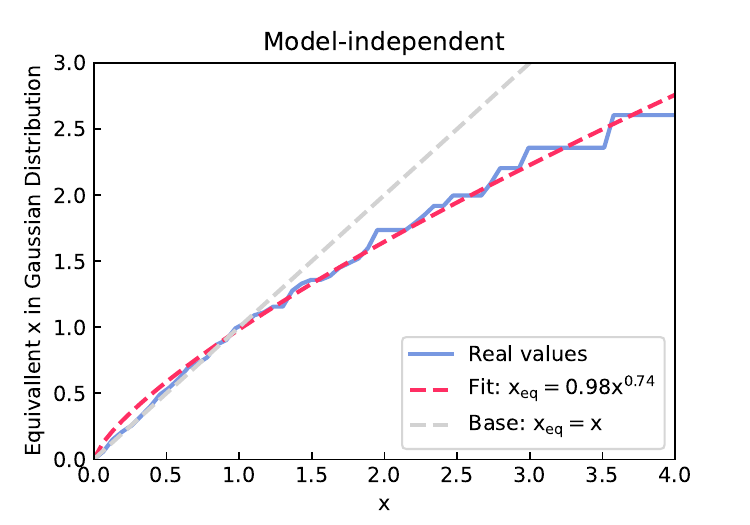}
	\includegraphics[width=0.49\textwidth]{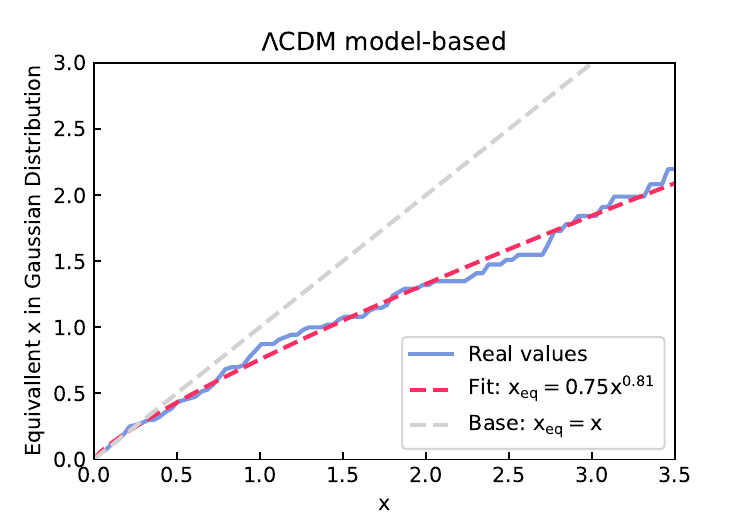}
	\caption{Equivalent number of $\sigma$ of a Guassian distribution as a function of the number of measured deviation
  $\sigma$ with respect to the weighted average value of $H_0$.
	The three panels (from top to bottom) correspond to the cases of 216 complete measurements, 109 model-independent measurements, and 107 $\Lambda$CDM model-based measurements, respectively.
	The red dashed lines represent the best fits, and the gray dashed lines denote the baselines (i.e., $x_{\rm eq}=x$).
		\label{fig:xeq}
	}
\end{figure}

In \autoref{fig:Prob}, we display the frequency of deviations larger than $x \sigma$ from the weighted average values $\overline{H_{0}}$ derived using the complete, model-independent and $\Lambda$CDM model-based categories.
For comparison, the Gaussian cumulative distributions are also given, which enables us to infer the degree to which the sample deviates from the expected Gaussian error distribution with a certain probability.
For instance, if $P=0.1$, the error is approximately $\sim 1.6 \sigma$ in a Gaussian error distribution, while $\sim 2.5 \sigma$ in real samples, indicating an overestimated deviation of $\sim 0.9 \sigma$.

To describe this overestimated deviation quantitatively, we calculate the equivalent deviation $x_{\rm eq}$ between the probabilities of Gaussian distributions and the real frequency, which means that there is actually a $x_{\rm eq}\sigma$ deviation when the deviation is $x \sigma$ in the real frequency.
And we fit them with the power function, $x_{\rm eq}=ax^b$, where $a$ and $b$ are free parameters. As shown in \autoref{fig:xeq}, the equivalent deviations $x_{\rm eq}$ in three categories are
\begin{eqnarray}\label{eq:xeq}
	x_{\rm eq}=(0.719\pm 0.013)x^{0.879\pm 0.014} &\,& {\text{(Complete)}},\\
	x_{\rm eq}=(0.983\pm 0.012)x^{0.744\pm 0.011} &\,& {\text{(Model-independent)}},\\
	x_{\rm eq}=(0.750\pm 0.007)x^{0.818\pm 0.009} &\,& 
    {\text{($\Lambda$CDM model-based)}}.
\end{eqnarray}
The baselines (i.e., $x_{\rm eq}=x$; gray dashed lines) are also plotted in \autoref{fig:xeq}, which represent
the expected deviations of a Gaussian distribution without underestimating the error bar.
Among three categories, the category of model-independent measurements most closely aligns with the Gaussian distribution (especially when $x\leq 1.5$), while others exhibit significant deviations.

Considering the current 5$\sigma$ tension reported in \cite{Riess2022}, the practical equivalent tension calculated using \autoref{eq:xeq} is $(2.96\pm 0.12) \,\sigma \sim 3\sigma$.
However, in order to calibrate $x_{\rm eq}$ and apply it to present-day Hubble tension values,
we should use data before the Hubble tension, because the calibration should be independent of the data whose tension we
want to test.
If we adopt the function of \cite{Martin2022} with data of 1976-2019, the equivalent tension is reduced to $2.25 \sigma$ ($x_{\rm eq}=0.83x^{0.62}$), which is a more accurate estimation.
It also indicates that the Hubble tension is stronger in the past decade, probably due to the improvement in measurement accuracy and the increase in relevant data.
Although these findings suggest an increase in the Hubble tension, underestimation of errors is still common.


\section{The Effects of Correlated Data}
\label{Sec:5}

In a large number of $H_0$ measurements, a substantial portion of the observational data is reused, which makes the measurements correlated, but we ignore the correlation between them.
As we mentioned $\chi^2$ in \autoref{chi2}, we treat the $H_0$ measurements as independent variables, and sum them
without considering the covariance matrix.
It is important to reiterate that the data we are investigating is not entirely independent, which means that our analysis may be biased. Therefore, it is necessary to account for the effects of correlated data.
In this section, we eliminate 64 correlated data, leaving 152 data (71 model-independent measurements and 81 $\Lambda$CDM-based measurements).
The removal criteria are that: a) if the measurements are obtained from the same data, we remove the earlier one; b) if one person measures $H_0$ using data A and data B, and the other person uses only data B, we discard the result that uses only data B.
We implement the same analysis and find minimal changes in all statistical characteristics as presented in \autoref{Tab4}.

\begin{table*}
	\centering
	\caption{The results of dip tests, weighted averages and $\chi^2$ fittings in three categories after removing some of the correlated measurements.
    The $Q$ values measure the statistical significance for $\chi^2$ fittings, and the $p$ values measure the significance of the unimodality for dip tests.}
	\label{Tab4}
	\renewcommand{\arraystretch}{1.2}
	\begin{tabular}{cccccc}
		\hline
		 & Number & $\overline{H_{0}} \; (\sigma)$ & $\chi^2$ & $Q$ & $p$ \\
		 & & $({\rm km\;s^{-1}\;Mpc^{-1}})$ & & \\
		\hline
		Complete & 152 & $69.25 \pm 0.13$ & 357.49 & $8.68 \times 10^{-19}$  & 0.12 \\
		Model-independent & 71 & $70.32 \pm 0.25$ & 123.61 & $8.20 \times 10^{-5}$ & 0.62 \\
		$\Lambda$CDM model-based & 81 & $68.99 \pm 0.15$ & 193.89 & $1.91 \times 10^{-11}$ & 0.13 \\
		\hline
  		$H_0<71 \; {\rm km\;s^{-1}\;Mpc^{-1}}$ & 85 & $68.84 \pm 0.09$ & 81.13 & 0.57 & 0.69 \\
		$H_0\ge 71 \; {\rm km\;s^{-1}\;Mpc^{-1}}$ & 67 & $73.65 \pm 0.15$ & 18.35 & $\sim 1.00$ & 0.98 \\
		\hline
	\end{tabular}
\end{table*}

We also examine the statistical properties of double peaks by categorizing the 152 $H_0$ measurements into two blocks: $H_0<71\;{\rm km\;s^{-1}\;Mpc^{-1}}$ (85 data) and $H_0\ge 71\;{\rm km\;s^{-1}\;Mpc^{-1}}$ (67 data).
The calculated results of the dip test are $p=0.12$ (complete), $p=0.69$ ($H_0<71\;{\rm km\;s^{-1}\;Mpc^{-1}}$) and $p=0.98$ ($H_0\ge 71\;{\rm km\;s^{-1}\;Mpc^{-1}}$),  respectively (see \autoref{Tab4}).
The bimodality is alleviated  but its $p$ value is far less significant than that of the two $H_0$ blocks.
Additionally, the results of $\chi^2$ fittings indicate two $H_0$ blocks still satisfy the Gaussian distributions ($Q \sim 1$).

However, there are still some correlated data in the dataset containing the 152 $H_0$ measurements, because it is not feasible to eliminate them using a fair approach.
For example, we can not exclude those $H_0$ measurements obtained by the same observational data but with different methods.
When we encounter the $H_0$ measurements separately obtained by the combined data A+B and the combined data B+C, although they are correlated, we also can not remove either of them.
In any case, if we had a random selection of duplicated points, the reduced $\chi ^2$ is similar, although the effective number of degrees of freedom is lower than the number assuming independence. Probability $Q$ might be larger
when covariance terms are taken into account, but still very low when $Q\ll 1$.


\section{{\bf Discussion and conclusions}}\label{Sec:Conclusion}

A statistical analysis of 163 measurements of the Hubble--Lema\^itre constant $H_0$ between 1976 and 2019 was performed by \cite{Martin2022}, indicating a potential underestimation of the statistical error bars or an inadequate consideration of the systematic errors.
Over the past decade, however, measurements of $H_0$ have increased in number and precision, making the Hubble tension increasingly important in a way that previous measurements did not.
Therefore, we compile a catalogue of 216 $H_0$ measured values from the years 2012-2022, including 109 model-independent measurements and 107 $\Lambda$CDM model-based measurements, to investigate the potential tension and biases over the last 11 years.

We find a significant bimodal distribution in the 216 $H_0$ measurements, corresponding to the results from the local distance ladder ($H_{0}=73.04 \pm 1.04 \; {\rm km\;s^{-1}\;Mpc^{-1}}$; \citealt{Riess2022}) and CMB observations ($H_{0}=67.4 \pm 0.5 \; {\rm km\;s^{-1}\;Mpc^{-1}}$; \citealt{Planck2020}), which has not been reported in previous statistical studies yet.
In the subsamples of 109 model-independent measurements and 107 $\Lambda$CDM model-based measurements, the bimodal distributions still exist.
We calculate the weighted averages $\overline{H_0}$ and the probabilities $Q$ for complete, model-independent and $\Lambda$CDM model-based measurements, yielding $\overline{H_0}=69.35\pm 0.12$ ($Q=8.85\times 10^{-27}$), $\overline{H_0}=70.82\pm 0.22$ ($Q=1.23\times 10^{-5}$) and $\overline{H_0}=68.94\pm 0.13$ ($Q=4.36\times 10^{-12}$) ${\rm km\;s^{-1}\;Mpc^{-1}}$, respectively.
Such low $Q$ values ($Q\ll 0.05$) indicate that they are lack of statistical significance.

The above results, for instance in \autoref{fig:Prob}, show clearly that
the deviations of the results with respect to the average $\overline{H_0}$ are far larger than expected from their error
bars if they follow a Gaussian distribution.
The underestimated 5$\sigma$ tension may actually be 3$\sigma$ when we recalibrate the frequency of deviations
with respect to the average, which is still a significant tension.
However, this recalibration should be independent of the data whose tension we
want to test. If we adopt the analysis of data of 1976-2019 \citep{Martin2022}, the equivalent tension is
reduced to $2.25\sigma $.

In addition, the separation of the data into two blocks with $H_0<71$ and $H_0\ge 71$ km s$^{-1}$ Mpc$^{-1}$ (Sect.
\ref{subc.71}) finds values of $Q\sim 1$, indicating compatibility with a Gaussian distribution for each of them without
removing any outliers. It points out that the possible underestimation of errors is related to own methodology in these
two groups of measurements.

At $H_0<71$ km s$^{-1}$ Mpc$^{-1}$, recent measurements of $H_0$ with low error bars are dominated by CMB measurements.
These values are subject to the errors in the cosmological interpretation of CMB with $\Lambda $CDM, and it is subject
to the many anomalies still pending to be solved in CMB anisotropies \citep{Sch16}.
Moreover, Galactic foregrounds are not perfectly removed \citep{Lop07,Axe15,Cre21}, and these are an important
source of uncertainties.

At $H_0\ge 71$ km s$^{-1}$ Mpc$^{-1}$, recent measurements of $H_0$ with low error bars are dominated by SN Ia teams, using
calibration with Cepheids.
Intrinsic scatters in SN Ia measurements are poorly understood \citep{Woj22}. The usual results are based on the assumption that there is only one hidden (latent) variable behind this scatter, namely the absolute magnitude.
With this assumption, one can claim that the error in the distance scale (and consequently in the $H_0$ measurement) can be reduced by increasing the number of observed supernovae. The problem is that the actual space of latent variables behind the intrinsic scatter is much larger: dust extinction in SN Ia depending on the type of host galaxies \citep{Mel23},
variations of the intrinsic luminosity of SN Ia with the age of the host galaxies \citep{Lee20}, etc.
Ignoring all these latent variables can only lead to underestimated errors and possible biases.
As a matter of fact, measurements showing a decrease of $H_0$ with $z$ in SN Ia measurements \citep{Jia23}
might indicate precisely the presence of these systematic biases rather than new physics or new cosmology.
We must also
bear in mind that the value of $H_0$ is determined without knowing on which scales the radial motion of galaxies and clusters of galaxies relative to us is completely
dominated by the Hubble--Lema\^itre flow. The homogeneity scale may be much larger than expected \citep{Syl11}, thus giving important net velocity flows on large scales that are incorrectly attributed to cosmological
redshifts.

In conclusion, our statistical analysis indicates that the underestimation of error bars for $H_0$ remains prevalent over the past decade, dominated by systematic errors in the methodologies of CMB and SN Ia analyses which make the tension become increasingly evident.

\section*{Acknowledgements}
We thank the anonymous referee for useful comments and suggestions.
This work is partially supported by the National Natural Science Foundation of China
(grant Nos. 12373053, 12321003, and 12041306), the Key Research Program of Frontier Sciences (grant No. ZDBS-LY-7014)
of Chinese Academy of Sciences, the Natural Science Foundation of Jiangsu Province (grant No. BK20221562),
and the Young Elite Scientists Sponsorship Program of Jiangsu Association for Science and Technology.
M.L.C. is supported by Chinese Academy of Sciences President's International Fellowship Initiative
(grant No. 2023VMB0001).

\section*{Data Availability}
The data underlying this article are available in the article and the cite references.



\bibliographystyle{mnras}
\bibliography{ref} 



\appendix

\section{Hubble--Lema\^itre Constant Catalogue}
\label{Catalogue}


\onecolumn

\renewcommand{\arraystretch}{1.1}
\begin{landscape}
	\begin{longtable}{ccc|ccc}
	\caption{216 measurements of the Hubble--Lema\^itre constant $H_0$ between the years 2012 and 2022.}
	\label{TabB1} \\
		\hline
		\multicolumn{3}{c}{Model-independent}  & \multicolumn{3}{c}{$\Lambda$CDM model-based} \\
            \hline
		 Year & $H_0$ $(\sigma)$ & References & Year & $H_0$ $(\sigma)$ & References \\
		  & $({\rm km\;s^{-1}\;Mpc^{-1}})$ & & & $({\rm km\;s^{-1}\;Mpc^{-1}})$ & \\
		\hline
2012 & $62.30\pm 6.30$   & Tammann and Reindl, 2012,   Ap\&SS, 341, 3.                             & 2012  & $69.10\pm 1.40$   & Campanelli et al., 2012, EPJC,   72, 2218.                                        \\
2012 & $65.00\pm 6.20$   & Lee and Jang, 2012, ApJL, 760,   L14.                                   & 2012  & $69.80\pm 1.20$   & Mehta et al., 2012, MNRAS, 427,   2168.                                           \\
2012 & $74.30\pm 2.10$   & Freedman et al. 2012, ApJ, 758,   24.                                   & 2012  & $76.50\pm 3.34$   & Holanda et al., 2012, GReGr, 44,   501.                                           \\
2012 & $74.30\pm 6.00$   & Ch\'{a}vez et al., 2012, MNRAS,   425, L57.            & 2013  & $69.00\pm 10.00$  & Sereno and Paraficz, 2014,   MNRAS, 437, 600.                                     \\
2012 & $75.90\pm 3.80$   & Courtois and Tully, 2012, ApJ,   749, 174.                              & 2013  & $70.00\pm 2.20$   & Hinshaw et al., 2013, ApJS, 208,   19.                                            \\
2013 & $64.00\pm 3.60$   & Tammann and Reindl, 2013,   A\&A, 549, A136.                            & 2013  & $71.80\pm 15.60$  & Dom\'{i}nguez and Prada, 2013,   ApJL, 771, L34. \\
2013 & $67.00\pm 3.20$   & Colless et al., 2013, IAUS, 289,   319.                                 & 2014  & $67.30\pm 1.20$   & Planck Collaboration et al.,   2014, A\&A, 571, A16.                              \\
2013 & $68.00\pm 9.00$   & Kuo et al., 2013, ApJ, 767,   155.                                      & 2014  & $68.50\pm 3.50$   & Verde et al., 2014, PDU, 5, 307.                                                  \\
2013 & $68.00\pm 4.80$   & Braatz et al., 2013, IAUS, 289,   255.                                  & 2014  & $74.10\pm 2.20$   & Lima and Cunha, 2014, ApJL, 781,   L38.                                           \\
2013 & $68.40\pm 6.30$   & Lee and Jang, 2013, ApJ, 773,   13.                                     & 2015  & $68.11\pm 0.86$   & Cheng and Huang, 2015, SCPMA,   58, 5684.                                         \\
2013 & $68.90\pm 7.10$   & Reid et al. 2013, ApJ, 767,   154.                                      & 2015  & $73.90\pm 10.05$  & Wei et al., 2015, MNRAS, 447,   479.                                              \\
2013 & $72.00\pm 3.00$   & Humphreys et al., 2013, ApJ,   775, 13.                                 & 2015  & $94.30\pm 34.25$  & Wei et al., 2015, AJ, 150, 35.                                                    \\
2013 & $72.40\pm 6.00$   & Di Benedetto, 2013, MNRAS, 430,   546.                                  & 2016  & $64.00\pm 12.50$  & Veropalumbo et al., 2016, MNRAS,   458, 1909.                                     \\
2013 & $74.00\pm 4.00$   & Sorce et al., 2013, ApJ, 765,   94.                                     & 2016  & $67.80\pm 0.90$   & Planck Collaboration et al.,   2016, A\&A, 594, A13.                              \\
2013 & $76.00\pm 1.90$   & Fiorentino et al., 2013, MNRAS,   434, 2866.                            & 2016  & $70.10\pm 0.34$   & Ichiki et al., 2016, PhRvD, 93,   023529.                                         \\
2014 & $72.00\pm 13.00$  & Jang and Lee, 2014, ApJ, 792, 52.                                       & 2017  & $66.40\pm 4.75$   & Melnick et al., 2017, A\&A,   599, A76.                                           \\
2014 & $75.20\pm 3.30$   & Sorce et al., 2014, MNRAS, 444,   527.                                  & 2017  & $67.60\pm 0.50$   & Alam et al., 2017, MNRAS, 470,   2617.                                            \\
2015 & $68.80\pm 3.30$   & Rigault et al., 2015, ApJ, 802,   20.                                   & 2017  & $67.60\pm 7.60$   & Cao et al., 2017, A\&A, 606,   A15.                                               \\
2015 & $69.80\pm 3.90$   & Jang and Lee, 2015, ApJ, 807,   133.                                    & 2017  & $67.87\pm 0.46$   & Chavanis and Kumar, 2017, JCAP,   2017, 018.                                      \\
2015 & $70.60\pm 2.60$   & Rigault et al., 2015, ApJ, 802,   20.                                   & 2017  & $69.13\pm 2.34$   & Wang et al., 2017, ApJ, 849, 84.                                                  \\
2016 & $71.90\pm 2.70$   & Bonvin et al. 2017, MNRAS, 465,   4914.                                 & 2017  & $70.10\pm 0.80$   & Wang and Meng, 2017, PDU, 18,   30.                                               \\
2016 & $73.24\pm 1.74$   & Riess et al. 2016, ApJ, 826, 56.                                        & 2017  & $71.75\pm 3.04$   & Wang et al., 2017, ApJ, 849, 84.                                                  \\
2016 & $73.75\pm 2.11$   & Cardona et al., 2017, JCAP,   2017, 056.                                & 2017  & $74.60\pm 2.10$   & Wang and Meng, 2017, PDU, 18,   30.                                               \\
2017 & $66.83\pm 2.21$   & Wu et al., 2017, Frontiers of   Physics, 12, 129801.                    & 2018  & $64.00\pm 10.00$  & Vega-Ferrero et al., 2018, ApJL,   853, L31.                                      \\
2017 & $67.38\pm 4.72$   & Wang and Meng, 2017, SCPMA, 60,   110411.                               & 2018  & $66.98\pm 1.18$   & Addison et al., 2018, ApJ, 853,   119.                                            \\
2017 & $68.30\pm 2.65$   & Chen et al., 2017, ApJ, 835,   86.                                      & 2018  & $68.00\pm 2.20$   & da Silva and Cavalcanti, 2018,   BrJPh, 48, 521.                                  \\
2017 & $69.13\pm 0.24$   & Huang and Huang, 2017, IJMPD,   26, 1750129.                            & 2018  & $69.40\pm 0.60$   & Burenin, 2018, AstL, 44, 653.                                                     \\
2017 & $69.30\pm 4.20$   & Braatz et al., 2018, IAUS, 336,   86.                                   & 2018  & $72.80\pm 4.20$   & Grillo et al., 2018, ApJ, 860,   94.                                              \\
2017 & $70.00\pm 10$  & Abbott et al., 2017, Nature,   551, 85.                                 & 2018  & $73.50\pm 4.65$   & Grillo et al., 2018, ApJ, 860,   94.                                              \\
2017 & $71.17\pm 3.53$   & Jang and Lee, 2017, ApJ, 836,   74.                                     & 2019  & $64.90\pm 4.45$   & Zeng and Yan, 2019, ApJ, 882,   87.                                               \\
2017 & $72.50\pm 3.87$   & Zhang et al., 2017, MNRAS, 471,   2254.                                 & 2019  & $66.60\pm 1.60$   & Dom\'{i}nguez et al., 2019,   ApJ, 885, 137.     \\
2017 & $73.10\pm 5.85$   & Wong et al., 2017, MNRAS, 465,   4896.                                  & 2019  & $67.00\pm 3.00$   & Kozmanyan et al., 2019,   A\&A, 621, A34.                                         \\
2018 & $71.00\pm 4.90$   & Fern\'{a}ndez Arenas et al.,   2018, MNRAS, 474, 1250. & 2019  & $67.40\pm 6.10$   & Dom\'{i}nguez et al., 2019,   ApJ, 885, 137.     \\
2018 & $71.90\pm 7.10$   & Cantiello et al., 2018, ApJL,   854, L31.                               & 2019  & $67.60\pm 1.10$   & Cuceu et al., 2019, JCAP, 2019,   044.                                            \\
2018 & $72.70\pm 2.10$   & Burns et al., 2018, ApJ, 869,   56.                                     & 2019  & $67.78\pm 1.54$   & Zhang et al., 2019, MNRAS, 483,   1655.                                           \\
2018 & $72.80\pm 4.30$   & Dhawan et al., 2018, A\&A,   609, A72.                                  & 2019  & $68.36\pm 0.53$   & Zhang and Huang, 2019, CoTPh,   71, 826.                                          \\
\hline

2018 & $73.15\pm 1.78$   & Feeney et al., 2018, MNRAS, 476,   3861.                                & 2019  & $68.44\pm 0.70$   & Ryane et al., 2019, MNRAS, 488,   3844.                                           \\
2018 & $73.20\pm 2.30$   & Burns et al., 2018, ApJ, 869,   56.                                     & 2019  & $68.80\pm 5.25$   & Birrer et al., 2019, MNRAS, 484,   4726.                                          \\
2018 & $73.30\pm 1.70$   & Follin and Knox, 2018, MNRAS,   477, 4534.                              & 2019  & $69.00\pm 1.70$   & Park and Ratra, 2019,   Ap\&SS, 364, 134.                                         \\
2018 & $73.48\pm 1.66$   & Riess et al., 2018, ApJ, 855,   136.                                    & 2019  & $69.30\pm 2.70$   & Jimenez et al., 2019, JCAP,   2019, 043.                                          \\
2019 & $69.80\pm 2.50$   & Freedman et al., 2019, ApJ, 882,   34.                                  & 2019  & $69.40\pm 2.00$   & Li et al., 2019, CoTPh, 71, 421.                                                  \\
2019 & $70.30\pm 5.15$   & Hotokezaka et al., 2019, NatAs,   3, 940.                               & 2019  & $71.00\pm 2.80$   & Jimenez et al., 2019, JCAP,   2019, 043.                                          \\
2019 & $71.10\pm 1.90$   & Reid et al., 2019, ApJL, 886,   L27.                                    & 2019  & $72.90\pm 2.20$   & Taubenberger et al., 2019,   A\&A, 628, L7.                                       \\
2019 & $72.20\pm 2.10$   & Liao et al., 2019, ApJL, 886,   L23.                                    & 2019  & $73.10\pm 2.15$   & Taubenberger et al., 2019,   A\&A, 628, L7.                                       \\
2019 & $72.40\pm 2.00$   & Yuan et al., 2019, ApJ, 886, 61.                                        & 2019  & $74.20\pm 2.95$   & Collett et al., 2019, PhRvL,   123, 231102.                                       \\
2019 & $73.50\pm 1.40$   & Reid et al., 2019, ApJL, 886,   L27.                                    & 2019  & $76.80\pm 2.60$   & Chen et al., 2019, MNRAS, 490,                                                    \\
2019 & $74.03\pm 1.42$   & Riess et al., 2019, ApJ, 876,   85.                                     & 2019  & $82.40\pm 8.35$   & Jee et al., 2019, Science, 365,   1134.                                           \\
2019 & $76.00\pm 16.00$  & Fishbach et al., 2019, ApJL,   871, L13.                                & 2020  & $65.90\pm 1.50$   & Holanda et al., 2020, JCAP,   2020, 053.                                          \\
2019 & $77.00\pm 27.50$  & Fishbach et al., 2019, ApJL,   871, L13.                                & 2020  & $67.40\pm 3.65$   & Birrer et al., 2020, A\&A,   643, A165.                                           \\
2019 & $78.00\pm 60.00$  & Soares-Santos et al., 2019,   ApJL, 876, L7.                            & 2020  & $67.40\pm 0.50$   & Planck Collaboration et al.,   2020, A\&A, 641, A6.                               \\
2020 & $64.80\pm 7.25$   & Howlett and Davi, 2020, MNRAS,   492, 3803.                             & 2020  & $67.60\pm 1.10$   & Aiola et al., 2020, JCAP, 2020,   047.                                            \\
2020 & $65.80\pm 5.90$   & Kim et al., 2020, ApJ, 905, 104.                                        & 2020  & $67.80\pm 0.70$   & Philcox et al., 2020, JCAP,   2020, 032.                                          \\
2020 & $66.20\pm 4.30$   & Dietrich et al., 2020, Science,   370, 1450.                            & 2020  & $67.90\pm 1.50$   & Aiola et al., 2020, JCAP, 2020,   047.                                            \\
2020 & $66.80\pm 11.30$  & Howlett and Davi, 2020, MNRAS,   492, 3803.                             & 2020  & $67.90\pm 1.10$   & Ivanov et al., 2020, JCAP, 2020,   042.                                           \\
2020 & $67.46\pm 4.75$   & Yang and Gong, 2020, JCAP, 2020,   059.                                 & 2020  & $67.95\pm 0.91$   & Zhang and Huang, 2020, SCPMA,   63, 290402.                                       \\
2020 & $68.60\pm 11.25$  & Nicolaou et al., 2020, MNRAS,   495, 90.                                & 2020  & $68.40\pm 0.58$   & Wang and Huang, 2020, JCAP,   2020, 045.                                          \\
2020 & $69.00\pm 21.50$  & Vasylyev and Filippenko, 2020,   ApJ, 902, 149.                         & 2020  & $68.50\pm 2.20$   & d'Amico et al., 2020, JCAP,   2020, 005.                                          \\
2020 & $69.60\pm 1.90$   & Freedman et al., 2020, ApJ, 891,   57.                                  & 2020  & $68.60\pm 1.10$   & Philcox et al., 2020, JCAP,   2020, 032.                                          \\
2020 & $70.00\pm 23.50$  & Vasylyev and Filippenko, 2020,   ApJ, 902, 149.                         & 2020  & $68.70\pm 1.50$   & Colas et al., 2020, JCAP, 2020,   001.                                            \\
2020 & $70.00\pm 9.00$   & Qi and Zhang, 2020, ChPhC, 44,   055101.                                & 2020  & $69.00\pm 1.20$   & Nadathur et al., 2020, PhRvL,   124, 221301.                                      \\
2020 & $70.40\pm 1.00$   & Antipova et al., 2020, AstBu, 75, 93.                                   & 2020  & $69.06\pm 0.63$   & Cao et al., 2020, MNRAS, 497,   3191.                                             \\
2020 & $71.50\pm 11.25$  & Wang et al., 2020, NatAs, 4,   517.                                     & 2020  & $69.60\pm 1.80$   & Pogosian et al., 2020, ApJL,   904, L17.                                          \\
2020 & $72.40\pm 7.60$   & Dhawan et al., 2020, ApJ, 888,   67.                                    & 2020  & $69.71\pm 1.28$   & Camarena and Marra, 2020, MNRAS,   495, 2630.                                     \\
2020 & $72.80\pm 3.80$   & Breuval et al., 2020, A\&A,   643, A115.                                & 2020  & $69.72\pm 1.63$   & Wang and Huang, 2020, JCAP,   2020, 045.                                          \\
2020 & $72.80\pm 1.65$   & Liao et al., 2020, ApJL, 895,   L29.                                    & 2020  & $71.60\pm 4.35$   & Rusu et al., 2020, MNRAS, 498,   1440.                                            \\
2020 & $73.00\pm 3.80$   & Breuval et al., 2020, A\&A,   643, A115.                                & 2020  & $72.30\pm 1.90$   & Nadathur et al., 2020, PhRvL,   124, 221301.                                      \\
2020 & $73.30\pm 4.00$   & Huang et al., 2020, ApJ, 889, 5.                                        & 2020  & $73.50\pm 5.30$   & Baxter and Sherwin, 2021, MNRAS,   501, 1823.                                     \\
2020 & $73.80\pm 6.05$   & Coughlin et al., 2020, NatCo,   11, 4129.                               & 2020  & $73.65\pm 2.11$   & Yang et al., 2020, MNRAS, 497,   L56.                                             \\
2020 & $73.90\pm 3.00$   & Pesce et al., 2020, ApJL, 891,   L1.                                    & 2020  & $74.00\pm 1.75$   & Millon et al., 2020, A\&A,   639, A101.                                           \\
2020 & $75.10\pm 3.80$   & Schombert et al., 2020, AJ, 160,   71.                                  & 2020  & $74.20\pm 1.60$   & Millon et al., 2020, A\&A,   639, A102.                                           \\
2020 & $75.10\pm 3.20$   & Kourkchi et al., 2020, ApJ, 902,   145.                                 & 2020  & $74.20\pm 2.85$   & Shajib et al., 2020, MNRAS, 494,   6072.                                          \\
2020 & $75.35\pm 1.68$   & Camarena and Marra, 2020, PhRvR, 2, 013028.                             & 2020  & $74.30\pm 1.90$   & Wei and Melia, 2020, ApJ, 897,   127.                                             \\
2020 & $75.80\pm 5.05$   & de Jaeger et al., 2020, MNRAS,   496, 3403.                             & 2020  & $74.36\pm 1.42$   & Camarena and Marra, 2020, MNRAS,   495, 2630.                                     \\
2020 & $79.00\pm 19.00$  & Coughlin et al., 2020, PhRvR, 2,   022006.                              & 2020  & $74.50\pm 5.85$   & Birrer et al., 2020, A\&A,   643, A165.                                           \\
2020 & $85.00\pm 19.50$  & Coughlin et al., 2020, PhRvR, 2,   022007.                              & 2020  & $75.32\pm 1.68$   & Camarena and Marra, 2020, MNRAS,   495, 2630.                                     \\
2020 & $109.00\pm 42.00$ & Coughlin et al., 2020, PhRvR, 2,   022008.                              & 2020  & $75.36\pm 1.68$   & Camarena and Marra, 2020, MNRAS,   495, 2630.                                     \\
2021 & $65.00\pm 22.50$  & GRAVITY Collaboration et al.,   2021, A\&A, 654, A85.                   & 2021  & $65.10\pm 4.20$   & Philcox et al., 2021, PhRvD,   103, 023538.                                       \\
\hline

2021 & $67.40\pm 4.80$   & Sun W., Jiao K., Zhang T.-J.,   2021, ApJ, 915, 123.                    & 2021  & $65.60\pm 4.45$   & Philcox et al., 2021, PhRvD,   103, 023538.                                       \\
2021 & $68.30\pm 4.55$   & Mukherjee et al., 2021,   A\&A, 646, A65.                               & 2021  & $67.30\pm 17.30$  & Wan et al., 2021, MNRAS, 504,   1062.                                             \\
2021 & $68.60\pm 11.75$  & Baklanov et al., 2021, ApJ, 907,   35.                                  & 2021  & $67.40\pm 0.90$   & Hall, 2021, MNRAS, 505, 4935.                                                     \\
2021 & $68.80\pm 35.60$  & Gayathri et al., 2021, ApJL,   908, L34.                                & 2021  & $67.49\pm 0.53$   & Balkenhol et al., 2021, PhRvD,   104, 083509.                                     \\
2021 & $69.00\pm 12.00$  & Abbott et al., 2021, ApJ, 909,   218.                                   & 2021  & $68.02\pm 1.82$   & Zhang X., Huang Q.-G., 2021,   PhRvD, 103, 043513.                                \\
2021 & $69.50\pm 4.00$   & Wang and Giannios, 2021, ApJ,   908, 200.                               & 2021  & $68.18\pm 0.79$   & Alam et al., 2021, PhRvD, 103,   083533.                                          \\
2021 & $69.80\pm 2.20$   & Freedman, 2021, ApJ, 919, 16.                                           & 2021  & $68.58\pm 1.76$   & Zhang X., Huang Q.-G., 2021,   PhRvD, 103, 043514.                                \\
2021 & $70.50\pm 5.75$   & Khetan et al., 2021, A\&A,   647, A72.                                  & 2021  & $68.63\pm 1.76$   & Zhang and Huang, 2021, PhRvD,   103, 043513.                                      \\
2021 & $71.80\pm 3.60$   & Denzel et al., 2021, MNRAS, 501,   784.                                 & 2021  & $68.80\pm 1.80$   & Cao et al., 2021, MNRAS, 504,   300.                                              \\
2021 & $72.10\pm 2.00$   & Soltis et al., 2021, ApJL, 908,   L5.                                   & 2021  & $68.80\pm 1.50$   & Dutcher et al., 2021, PhRvD,   104, 022003.                                       \\
2021 & $73.20\pm 1.30$   & Riess et al., 2021, ApJL, 908,   L6.                                    & 2021  & $68.90\pm 1.05$   & Ivanov, 2021, PhRvD, 104,   103514.                                               \\
2021 & $73.30\pm 3.10$   & Blakeslee et al., 2021, ApJ,   911, 65.                                 & 2021  & $69.30\pm 1.20$   & Cao et al., 2021, MNRAS, 501,   1520.                                             \\
2021 & $73.78\pm 0.84$   & Bonilla et al., 2021, EPJC, 81,   127.                                  & 2021  & $70.00\pm 2.70$   & Addison, 2021, ApJL, 912, L1.                                                     \\
2022 & $65.90\pm 2.95$   & Zhang et al., 2022, ApJ, 936,   21.                                     & 2021  & $70.60\pm 3.70$   & Philcox et al., 2021, PhRvD,   103, 023538.                                       \\
2022 & $68.10\pm 2.60$   & M\"{o}rtsell et al., 2022,   ApJ, 935, 58.             & 2021  & $72.40\pm 4.35$   & Addison, 2021, ApJL, 912, L1.                                                     \\
2022 & $68.10\pm 3.50$   & M\"{o}rtsell et al., 2022,   ApJ, 933, 212.            & 2021  & $73.10\pm 3.60$   & Addison, 2021, ApJL, 912, L1.                                                     \\
2022 & $68.80\pm 11.85$  & Gray et al., 2022, MNRAS, 512,   1127.                                  & 2021  & $73.60\pm 1.70$   & Qi et al., 2021, MNRAS, 503,   2179.                                              \\
2022 & $68.80\pm 3.65$   & Benndorf et al., 2022, EPJC, 82,   457.                                 & 2021  & $84.50\pm 6.40$   & Ivanov, 2021, PhRvD, 104,   103514.                                               \\
2022 & $71.50\pm 1.80$   & Anand et al., 2022, ApJ, 932,   16.                                     & 2022  & $62.30\pm 9.10$   & Hagstotz et al., 2022, MNRAS,   511, 662.                                         \\
2022 & $71.90\pm 2.20$   & M\"{o}rtsell et al., 2022,   ApJ, 933, 212.            & 2022  & $64.80\pm 2.35$   & Philcox et al., 2022, PhRvD,   106, 063530.                                       \\
2022 & $72.19\pm 5.74$   & Gallego-Cano et al., 2022,   A\&A, 666, A13.                            & 2022  & $67.10\pm 2.70$   & Philcox et al., 2022, PhRvD,   106, 063530.                                       \\
2022 & $72.24\pm 2.63$   & Liu et al., 2022, ApJ, 939, 37.                                         & 2022  & $68.19\pm 0.99$   & Zhang et al, 2022, JCAP, 2022,   036.                                             \\
2022 & $72.45\pm 1.99$   & Liu et al., 2022, ApJ, 939, 38.                                         & 2022  & $68.30\pm 0.45$   & Sch\"{o}neberg et al.,   2022, JCAP, 2022, 039.                  \\
2022 & $72.90\pm 2.30$   & Kenworthy et al., 2022, ApJ,   935, 83.                                 & 2022  & $68.81\pm 4.66$   & Wu et al., 2022, MNRAS, 515, L1.                                                  \\
2022 & $73.01\pm 0.99$   & Riess et al., 2022, ApJ, 938,   36.                                     & 2022  & $69.10\pm 0.14$   & Zhang and Cai, 2022, JCAP, 2022,   031.                                           \\
2022 & $73.04\pm 1.04$   & Riess et al., 2022, ApJL, 934,   L7.                                    & 2022  & $69.50\pm 3.25$   & Farren et al., 2022, PhRvD, 105,   063503.                                        \\
2022 & $73.15\pm 0.97$   & Riess et al., 2022, ApJ, 938,   36.                                     & 2022  & $69.70\pm 1.20$   & Cao and Ratra, 2022, MNRAS, 513,   5686.                                          \\
2022 & $73.20\pm 1.30$   & M\"{o}rtsell et al., 2022,   ApJ, 933, 212.            & 2022  & $69.70\pm 1.20$   & Cao et al., 2022, MNRAS, 509,   4745.                                             \\
2022 & $73.90\pm 1.80$   & M\"{o}rtsell et al., 2022,   ApJ, 933, 212.            & 2022  & $73.00\pm 10.00$  & James et al., 2022, MNRAS, 516,   4862.                                           \\
2022 & $75.50\pm 2.50$   & Kourkchi et al., 2022, MNRAS,   511, 6160.                              & 2022  & $73.42\pm 1.79$   & Liu et al., 2022, A\&A, 668,   A51.                                               \\
2022 & $76.70\pm 2.00$   & M\"{o}rtsell et al., 2022,   ApJ, 933, 212.            & &  & \\
2022 & $76.94\pm 6.40$   & Dhawan et al., 2022, ApJ, 934,   185.                                   &  &  &  \\
\hline
\end{longtable}
\begin{description}
\item {\it Note.} The asymmetrical error bars are averaged.
\end{description}
\end{landscape}


\bsp	
\label{lastpage}
\end{document}